\documentclass[aps,pra,superscriptaddress,twocolumn]{revtex4}

\renewcommand*{\[}{\begin{equation}}

\renewcommand*{\]}{\end{equation}}

\usepackage{amsmath}
\usepackage{graphicx}
\usepackage{color}

\usepackage{hyperref}

\begin{document}

\title{Enhanced high harmonic generation in semiconductors by the excitation with multi-color pulses}
\author{Xiaohong Song}

\affiliation{Research Center for Advanced Optics and Photoelectronics, Department of Physics, College of Science, Shantou University, Shantou, Guangdong 515063, China}
\affiliation{Key Laboratory of Intelligent Manufacturing Technology (Shantou University), Ministry of Education}


\author{Shidong Yang}
\affiliation{Research Center for Advanced Optics and Photoelectronics, Department of Physics, College of Science, Shantou
University, Shantou, Guangdong 515063, China}

\author{Ruixin Zuo}
\affiliation{Research Center for Advanced Optics and Photoelectronics, Department of Physics, College of Science, Shantou
University, Shantou, Guangdong 515063, China}

\author{Torsten Meier}
\affiliation{Department of Physics and Center for Optoelectronics and Photonics Paderborn (CeOPP), University of Paderborn,
Warburger Str. 100, D-33098 Paderborn, Germany}

\author{Weifeng Yang}
\email{wfyang@stu.edu.cn}
\affiliation{Research Center for Advanced Optics and Photoelectronics, Department of Physics, College of Science, Shantou University,
Shantou, Guangdong 515063, China}
\affiliation{Key Laboratory of Intelligent Manufacturing Technology (Shantou University), Ministry of Education}
\date{\today}

\begin{abstract}
We investigate high-order harmonic generation in ZnO driven by linearly polarized multi-color pulses. It is shown that the intensities of the harmonics in the plateau region can be enhanced by two to three orders of magnitude when driven by two- or three-color fields as compared with the single-color pulse excitation. By analyzing the time-dependent population in the conduction band as function of both the initial and the moving crystal momenta, we demonstrate that this remarkable enhancement originates from the intraband preacceleration of electrons from their initial momenta to the top of the valence band where interband excitation takes place. We show that this preacceleration strongly increases the population in the conduction band and correspondingly the intensities of high harmonics in the plateau region. Our results confirm the very recently proposed four-step model for high harmonic generation in semiconductors [Phys. Rev. Lett. \textbf{122},
193901 (2019)] and provide an effective way to enhance the intensities of harmonics in the plateau region which has significant implications for the generation of bright attosecond light sources from semiconductors.
\end{abstract}

\maketitle

\section{Introduction}
\label{intro}

High harmonic generation (HHG) from atoms and molecules has been intensively studied over the last three decades, which led to successful applications such as attosecond pulse generation, tracing ultrafast electron dynamics, and imaging molecular orbitals and structures \cite{Krausz2009,Corkum2007,Itatani2004}. The basic processes involved in HHG from gases can be described by a semiclassical three-step model: (i) the electron tunnels through the potential barrier due to the action of a strong laser field, (ii) the electron is accelerated away from the atom/molecule by the laser field, (iii) the electron returns to the ionic core and under suitable conditions recombines under emission of a high energy photon \cite{Corkum1993}. Several studies showed that the entire process depends sensitively on the waveform of the laser field. Guided by this simple model, various laser synthesis and shaping schemes have been proposed to control and optimize HHG as needed. For example, to achieve a high conversion efficiency, synthesized two- or three-color laser fields have been adopted such that electrons are born with a higher ionization rate or short-trajectory electrons are enhanced over long-trajectory ones to improve phase-matching \cite{Watanabe1994, Jin2014,Jin2019,Zhang2015}. To realize higher cutoff energies, an ideal waveform consisting of a linear ramp with a dc offset has been proposed to generate the maximal possible electron recollision energy for any given oscillation period \cite{Chipperfield2009}. To achieve isolated attosecond pulse, few-cycle polarization gating, waveform-controlled sub-1.5-cycle laser pulse, and multi-color waveform synthesis have been adopted to confine the emission within one half-cycle of the field \cite{solar2006,Goulielmakis2008,Sansone2006,Wirth2011,Zeng2007,Song2007}. Thanks to the advancement in optical parametric amplification and optical parametric chirped-pulse amplification technology, nowadays, it is possible to generate nearly arbitrary optical waveforms by synthesizing multi-color laser pulse in experiments, making the generated attosecond pulses a powerful tool for the real-time observation of ultrafast electronic dynamics in matter \cite{Krausz2009,Wirth2011,Huang2011,Burger2017,Schultze2010,Pazourek2015}.

Several recent experimental investigations have demonstrated that HHG can also be obtained from bulk solids, opening a new direction for more compact attosecond sources and on-chip ultrafast photonics \cite{Ghimire2011,Schubert2014,Hohenleutner2015,Luu2015,Vampa2015,Ndabashimiye2016,Liu2017,Vampa2017,Yu2018,Huttner2017}. However, up to date, the physical picture of the microscopic processes governing HHG from solids is still not fully established and remains a topic of intense debate. A three-step model in momentum space has been proposed for solid HHG \cite{Vampa2014,2Vampa2015,Wu2016PRA}: (i) an electron tunnels vertically from the valence to the conduction band so that an electron-hole pair is formed (interband transition), (ii) the laser subsequently accelerates the pair within the bands (intraband motion), (iii) the electron recombines with the hole and emits a harmonic photon with an energy given by the energy difference between the conduction and valence band states at the final momentum. Despite the similarities, there are remarkably difference between atomic and solid HHG. Firstly, in atomic HHG the ionization process (i.e., the first step) is not influenced by electron motion in the continuum state (i.e., the second step) since the electron starts from a bound state. However, in semiconductors the electrons occupy delocalized states in the valence and conduction bands both prior and after the excitations and it has been demonstrated that
the `vertical' interband transition and the intraband motion within the electronic band are not independent \cite{Golde2008}. The complex coupling between the two mechanisms affects not only the carrier injection from the valence into the conduction band, but also the motion of excited electron wavepacket in momentum space \cite{Wismer2016,Schlaepfer2018,Dejean2018,Song2019}. Secondly, unlike atomic HHG, the purely intraband oscillation represents an additional source for the HHG emission of solids \cite{Luu2015,Liu2017,Wu2016PRA}. Thirdly, different to the atomic case, in solids also the hole moves during the excitation process, and as a results the electron can collide with neighboring atomic sites of the periodic crystal potential \cite{You2017,Liu2017PRA}. Very recently, it has been demonstrated that contrary to atomic systems, the cutoff of the HHG from solids is strongly increased by exciting with a circularly polarized field. Moreover, the harmonic yield with large ellipticity is comparable to or even higher than that in a linearly polarized field, and to understand the underlying physics a new four-step model has been proposed \cite{LiLiang2019}.

In view of the abovementioned differences, it is relevant to analyze whether methods that were developed for controlling HHG in the gas-phase are also applicable to solid HHG.
In atomic systems, it has been shown that adding an additional third harmonic field with an intensity of only 10\% can enhance the HHG in the plateau region by an order of magnitude \cite{Watanabe1994}. Here, we show that the intensity of HHG in semiconductor ZnO can be increased by at least two orders by the same field ratio ($E_{3\omega}/E_{\omega}$) as that used in  atomic media.
We analyze this effect theoretically and demonstrate that the stronger enhancement of HHG in semiconductors is due to the preacceleration step, i.e., an intraband acceleration, which takes place prior to ionization, i.e., the interband excitation.
Consistent with this interpretation we find that the HHG can be increased further by adding an additional weak fifth harmonic field.
Altogether our findings show an effective way to enhance the intensities of harmonics in the plateau region
and our analysis confirms the recently proposed four-step model of HHG in solids \cite{LiLiang2019}.

This paper is organized as follows.
In Sec.~\ref{theory} we describe our theoretical approach, i.e., the semiconductor Bloch equations (SBE) for a two-band model, and introduce the two frames that we use in momentum space
to analyze and interpret our findings.
The results of our numerical simulations are presented and discussed in Sec.~\ref{res}.
Our main results are briefly summarized in Sec.~\ref{sum}.


\section{Theoretical approach}
\label{theory}

The HHG in semiconductors is analyzed using the semiconductor Bloch equations which include the coupled inter- and intraband dynamics and for a two-band model read \cite{Golde2008,Golde2011}:
\begin{equation}
\begin{split}
i\hbar\frac{\partial}{\partial{t}}p_{\mathbf{k}}=&(\varepsilon_{\mathbf{k}}^{e}+\varepsilon_{\mathbf{k}}^{h}-i\frac{\hbar}{T_{2}})p_{\mathbf{k}}-(1-n_{\mathbf{k}}^{e}-n_{\mathbf{k}}^{h})\textbf{d}_{\mathbf{k}}\cdot\textbf{E}(t)\\
&+ie\textbf{E}(t)\cdot\nabla_{\mathbf{k}}p_{\mathbf{k}},
\end{split}
\end{equation}
\begin{equation}
\hbar\frac{\partial}{\partial{t}}n_{\mathbf{k}}^{e(h)}=-2\mathrm{Im}[\textbf{d}_{\mathbf{k}}\cdot \textbf{E}(t)p_{\mathbf{k}}^{*}]+e\textbf{E}(t)\cdot\nabla_{\mathbf{k}}n_{\mathbf{k}}^{e(h)}.
\end{equation}
Here, $\varepsilon_{\mathbf{k}}^{e(h)}$ are the band energies of electrons (holes), $T_{2}$ is the dephasing time, $\textbf{d}_{\mathbf{k}}$ is the dipole matrix element, and $\textbf{E}(t)$ denotes the electric field. According to the Bloch acceleration theorem \cite{Bloch1929}, in the presence of a homogeneous electric field the electron's crystal momentum in a given band changes according to $\textbf{K}(t)=\textbf{k}-\textbf{A}(t)$, where $\textbf{A}(t)$ is the vector potential of the electric field $-d\textbf{A}/dt=\textbf{E}(t)$ and $\textbf{k}=\textbf{K}(t_{0})$ is the initial momentum.
As a result, the semiconductor Bloch equations can also be transformed to and solved in a moving moving frame $\textbf{K}(t)$ \cite{Vampa2014}:
\begin{equation}
\frac{\partial}{\partial{t}}p_{\mathbf{K}}=-\frac{p_{\mathbf{K}}}{T_{2}}-i \Omega(\mathbf{K}, t) (1-n_{\mathbf{K}}^{e}-n_{\mathbf{K}}^{h}) e^{-i S(\mathbf{K}, t)},
\end{equation}
\begin{equation}
\frac{\partial}{\partial{t}}n_{\mathbf{K}}^{e(h)}=i  \Omega^{*}(\mathbf{K}, t) p_{\mathbf{K}} e^{i S(\mathbf{K}, t)}+\mathrm{c.c.}.
\end{equation}
Here, $\Omega(\mathbf{K}, t)=\mathbf{E}(t) \cdot \mathbf{d}[\mathbf{K}+\mathbf{A}(t)]/\hbar$ is the Rabi frequency, $S(\mathbf{K},t)=\hbar^{-1}\int_{-\infty}^{t}\varepsilon_{\mathrm{g}}\left[\mathbf{K}+\mathbf{A}\left(t^{\prime}\right)\right] d t^{\prime}$ is the classical action, and $\varepsilon_{g}[\mathbf{k}]=\varepsilon_{\mathbf{k}}^{e}+\varepsilon_{\mathbf{k}}^{h}$ is the k-dependent transition energy between the valence and conduction bands.
Clearly, the two frames are connected by a transformation and thus provide the same results. It is, however, very instructive to compare the time-dependent dynamics in the two frames since this provides a clear picture of the electrons' dynamics as we demonstrate below.

In this work, we focus on HHG of ZnO by linearly-polarized single-color ($\omega$),  two-color ($\omega/3\omega$), and  three-color ($\omega/3\omega/5\omega$) fields. The intensity of the electric field is chosen so that the excitation to the higher conduction bands is negligible and therefore a two-band model is adopted. This has been checked by simulations performed with three-bands where we obtained nearly the same results as for the two-band model.
The band structure parameters of ZnO are taken from Ref.~\cite{Vampa2015PRB}.
The electric field is written as:
\begin{equation}
\begin{split}
E(t)=&f(t)[E_{\omega}cos(\omega_{0}t)+E_{3\omega}cos(3\omega_{0}t+\phi_{1})\\
&+E_{5\omega}cos(5\omega_{0}t+\phi_{2})],
\end{split}
\end{equation}
where $f(t)=\mathrm{exp}[-2\mathrm{ln}2(t-t_{0})^2/\tau_{p}^{2}]$ is the pulse envelope with the full width at half maximum (FWHM) of the pulse $\tau_{p}=35 \mathrm{fs}$,
$E_{\omega},E_{3\omega},E_{5\omega}$ represent the amplitudes, $\omega_{0}$ is the fundamental frequency, and $\phi_{1,2}$ denote the relative phase between the fundamental and third or fifth harmonic, respectively.
In the following, $E_{3\omega}=E_{5\omega}=0$ for the one-color pulse and $E_{5\omega}=0$ for the two-color pulse.
Since the electric field is linearly polarized along the $\Gamma$-M direction of the Brillouin zone, the Bloch equations are solved for a one-dimensional model taking the $\Gamma$-M band structure into account.

\begin{figure}[htb]
\includegraphics[width=1.05\columnwidth]{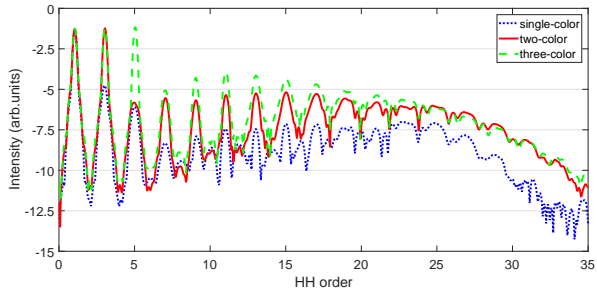}
\caption{Simulated HHG spectra for ZnO generated by single-color (dotted blue line), two-color (solid red line), and three-color (dashed green line) pulses, respectively, with fundamental wavelength $\lambda_0 = 3.2 \mathrm{{\mu}m}$ and an amplitude of the fundamental pulse $E_{\omega} = 0.003\mathrm{a.u.}$ on a logarithmic scale. Further parameters are given in the text.}
\label{Figure 1}
\end{figure}

\begin{figure}[htb]
\includegraphics[width=1.05\columnwidth]{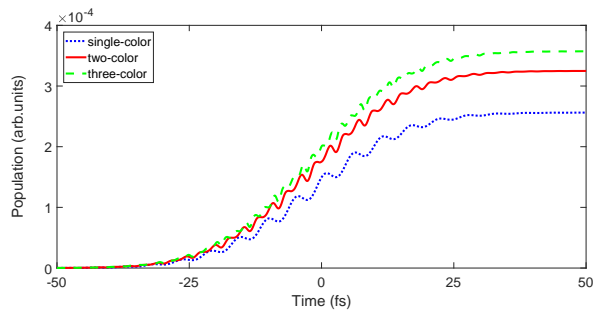}
\caption{The total time-dependent population in the conduction band excited by single-color (dotted blue line), two-color (solid red line), and three-color (dashed green line) pulses, respectively, using the same parameters as in Fig.~1.}
\label{Figure 2}
\end{figure}

\begin{figure*}[htb]
\centering
\includegraphics[width=2\columnwidth]{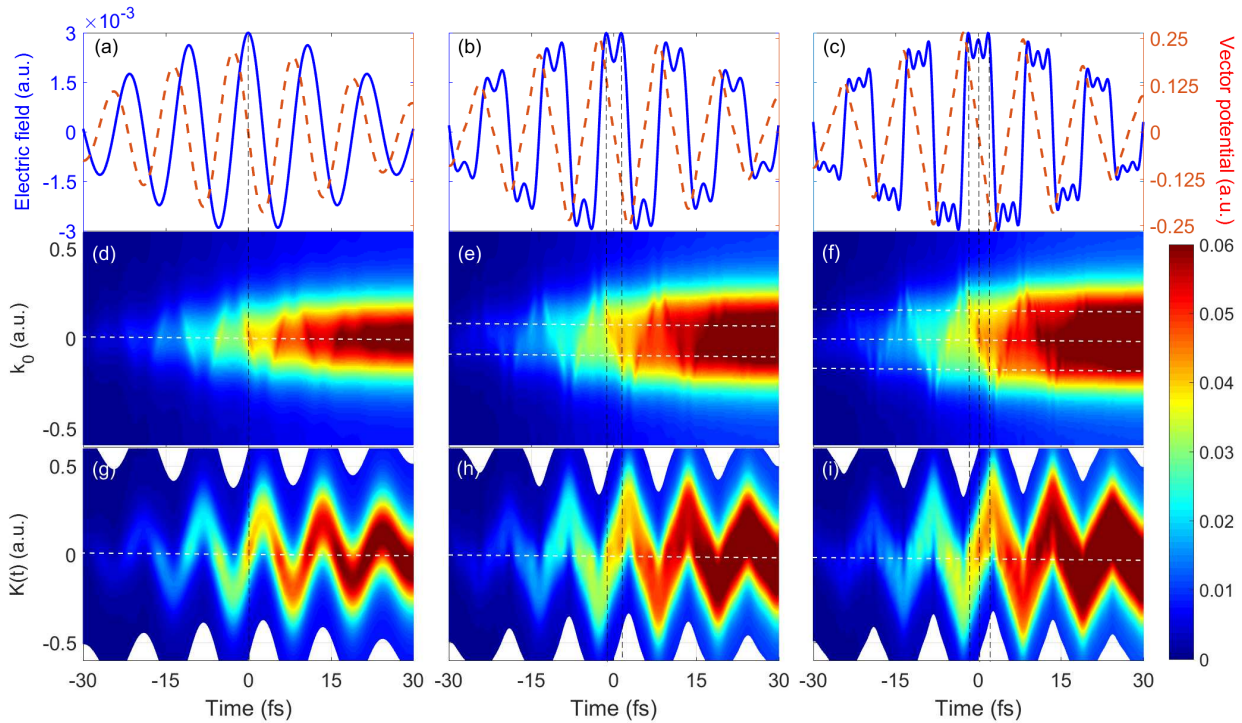}
\caption{(a), (b), and (c): The electric field (solid blue line) and vector potential (dashed red line) for single-color, two-color, and three-color excitation, respectively. (d), (e), and (f): The corresponding time-dependent conduction band population in k-space. (g), (h), and (i): The time-dependent conduction band population in the moving K-space frame. The vertical black dashed lines represent the temporal positions of excitation bursts that occur at the maxima of the electric field. The horizontal white dashed lines indicate the momenta at which the strongest excitation takes place. In the k-space these values correspond to the values of the vector potential at the maxima of the electric field.}
\label{Figure 3}
\end{figure*}

\section{Results and discussion}
\label{res}

Figure~1 shows HHG spectra generated by three different optical fields: the single fundamental one-color pulse (blue dotted line) with intensity of $3.16 \times 10^{11} \mathrm{W} / \mathrm{cm}^{2}$, the two-color pulse with field ratio of $E_{3\omega}/E_{\omega}=0.3$ and a relative phase $\phi_{1}=\pi$ (red solid line), and the three-color pulse with $E_{3\omega}/E_{\omega}=0.3$, $E_{5\omega}/E_{\omega}=0.2$ and $\phi_{1}=\pi$, $\phi_{2}=0$ (green dashed line). In Ar atoms, it has been demonstrated that the intensities of HHG in the plateau region are enhanced by an order of  magnitude by a field ratio of $E_{3\omega}/E_{\omega}=0.3$ \cite{Watanabe1994}. Here we find that compared with the single-color pulse case, the harmonic intensities in the plateau region are enhanced for the two-color pulse by more than two orders of magnitude (red solid line). It should be noted that though the intensity and the frequency of the fundamental are different to those in Ref.~\cite{Watanabe1994} due to the different medium, the same field ratio is used as in the atomic case. When further adding an even weaker fifth harmonic field, the intensity of HHG is even three orders stronger than that for the single-color field (green dashed line). In Fig.~2, the $k$-integrated population in the conduction band are shown for these three different fields. Clearly, the populations for both the two-color and the three-color fields are much higher than that of the single-color field, and the population in the three-color field is the strongest, which is consistent with the HHG shown in Fig.~1.\par
In order to further analyze the physics behind the strongly enhanced HHG for the two-color and three-color fields, we present in Fig.~3 the time-dependent population in the conduction band as function of initial momentum $\textbf{k}$ (see Figs.~3(d)-(f)) and as function of the moving momentum $\textbf{K}(t)$ (see Figs.~3(g)-(i)), respectively. Figs.~3(a)-3(c) show the electric field (blue solid lines) and the corresponding vector potential (red dashed lines). For the single-color field case, in the initial momentum $\textbf{k}$ frame the excitation, i.e., the temporal positions when the population increases rapidly, occurs mainly around $\textbf{k}=0$, i.e., near the $\Gamma$ point which is when the electric field reaches its extrema and the vector potential equals to 0 (as indicated by the white and black dotted lines). The same situation is seen also in the moving $\textbf{K}$ frame (see Fig.~3(g)). The only difference is that in the moving frame the population is distributed over a larger region in $\textbf{K}$-space due to the intraband acceleration. This confirms the three-step model for solids, i.e., the electrons are firstly excited to the conduction band near the $\Gamma$ point, then both electrons and holes move in the corresponding bands under the action of the laser field according the acceleration theorem and subsequently emit high-order harmonic photons.

For the case of the two-color field with $\phi_{1}=\pi$, the electric field amplitude has two maxima within one half-cycle. Correspondingly, two excitation bursts are observed in both the initial $\textbf{k}$ and moving $\textbf{K}(t)$ momentum frames at the same times. In contrast to the single-color field case, these excitations do not occur at $\textbf{k}=0$ in the initial momentum frame but at $\textbf{k}\approx\pm0.11$ a.u. (as indicated by the white dotted lines). Whereas in the moving frame, these excitation bursts appear at $\textbf{K}(t)\approx0$ since the strongest transitions appear when the accelerated electrons pass through the $\Gamma$ point. It should be noticed that for the two-color field, the vector potential is not equal to 0 at the times where the electric field reaches its maxima. When we compare the relationship of the population bursts in the two different momentum frames, we find that it satisfies the acceleration theorem $\textbf{K}\approx\textbf{k}-\textbf{A}(t)$, which means that actually the excitation mainly occurs whenever $\textbf{K}(t)=0$. However, this clearly does not correspond to $\textbf{k}=0$ since $\textbf{k}$ and $\textbf{K}$ differ by the vector potential. Thus our findings confirm the very recently proposed four-step model of HHG in solids, i.e., before interband excitation (also be called "ionization") to the conduction band, the electron is accelerated for a while until it reaches to the top of the valence band \cite{LiLiang2019}. It has been demonstrated that the cutoff of HHG is strongly extended in a circularly polarized field due to this preacceleration process \cite{LiLiang2019}. Here we demonstrate that this preacceleration process has also consequences for linearly polarized two-color fields and results in a strong enhancement of the HHG and the population in the conduction band.

To further demonstrate relevance of the preacceleration process, we consider the three-color excitation with $\phi_{1}=\pi$ and $\phi_{2}=0$, and field ratios of $E_{3\omega}/E_{\omega}=0.3$ and $E_{5\omega}/E_{\omega}=0.2$. In this case, there are three subpeaks present within one half-cycle of the electric field. For the central half-cycle, the corresponding vector potential at these sub-peaks are $A_{1}=0.19$ a.u., $A_{2}=0$, and $A_{3}=-0.19$ a.u., respectively. In Fig.~3(f) three excitation bursts can be observed which are located at $\textbf{k}=\pm0.19$ a.u. and $\textbf{k}=0$ in the initial $\textbf{k}$ momentum frame. In the moving frame, all these excitation bursts move to around $\textbf{K}(t)\approx0$. The three-color excitation leads to a further increase of the population in the conduction band (see Fig.~2) and the intensity of the harmonics in the plateau region is enhanced by 3 orders of magnitude compared to the single-color case (see Fig.~1).

\section{Summary}
\label{sum}

In conclusion, we have investigated HHG from ZnO driven by single-color ($\omega$), two-color ($\omega-3\omega$), and three-color ($\omega-3\omega-5\omega$) fields. We found that the intensities of the harmonics in the plateau are substantially enhanced by about two orders of magnitude for the two-color excitation and three orders for the three-color excitation compared to the single-color case. By comparing the time-dependent population in the conduction band in the $\textbf{k}$ and the $\textbf{K}(t)$ frames, the relevance of the preacceleration step to $\textbf{K}\approx0$ prior to interband excitation is clearly demonstrated, which confirms the recently proposed four-step model for HHG of solids. We have shown that for linearly-polarized pulse excitation, together with the HHG also the population in the conduction band is significantly increased due to the preacceleration step. Our results not only present a practical method for enhancing the intensity of harmonics in the plateau region, but also pave the way toward a more complete understanding of the physical mechanism that govern HHG of solids.

\acknowledgments
The work was supported by the NNSF of China (Grant No. 11674209, No. 11774215), Department of Education of Guangdong Province (Grant No. 2018KCXTD011), High Level University Projects of Guangdong Province (Mathematics, Shantou University), Open Fund of the State Key Laboratory of High Field Laser Physics (SIOM), and the Deutsche Forschungsgemeinschaft (DFG, German Research Foundation) ¨C project number 231447078 ¨C TRR 142 (project A07).

\end{document}